\documentclass[journal]{IEEEtran}

\usepackage{graphicx}
\usepackage[dvipsnames]{xcolor}
\usepackage{amssymb}
\usepackage{amsmath}
\usepackage{optidef}
\usepackage{array}
\usepackage{algpseudocode}
\usepackage{multirow}
\usepackage{algorithm}
\usepackage{cite}
\usepackage{csquotes}
\usepackage{float}
\usepackage{todonotes}
\usepackage{soul}
\usepackage[hidelinks]{hyperref}
\usepackage{diagbox}
\usepackage[bottom]{footmisc}

\usepackage{lipsum}
\ifCLASSINFOpdf
\else
\fi
\begin{document}
\title{Swin Transformer-Based Dynamic Semantic Communication for Multi-User with Different Computing Capacity}

\author{Loc X. Nguyen, Ye Lin Tun,  Yan Kyaw    Tun,~\IEEEmembership{Member,~IEEE,}
        Minh N. H. Nguyen,~\IEEEmembership{Member,~IEEE,}
        Chaoning Zhang,~\IEEEmembership{Member,~IEEE,}
         Zhu Han,~\IEEEmembership{Fellow,~IEEE,}
         and Choong Seon Hong,~\IEEEmembership{Senior Member,~IEEE}% <-this % stops a space
\thanks{Loc X. Nguyen, Ye Lin Tun, Choong Seon Hong are with the Department of Computer Science and Engineering, Kyung Hee University, Yongin-si, Gyeonggi-do 17104, Rep. of Korea, e-mail: \{xuanloc088,yelintun,cshong\}@khu.ac.kr. }% <-this % stops a space
\thanks{Yan Kyaw Tun is with the Division of Network and Systems Engineering, School of Electrical Engineering and Computer Science, KTH Royal Institute of Technology, Brinellvägen 8, 114 28 Stockholm, Sweden e-mail:{\{yktun\}@kth.se}.}
\thanks{Minh N. H. Nguyen is with the Digital Science and Technology Institute, The University of Danang—Vietnam-Korea University of Information and Communication Technology, Da Nang 550000, Vietnam (e-mail:nhnminh@vku.udn.vn).}
\thanks{Chaoning Zhang is with the Department of Artificial Intelligence, School of Computing, Kyung Hee University, Yongin-si 17104, Republic of Korea (email:chaoningzhang1990@gmail.com).}

\thanks{Zhu Han is with the Electrical and Computer Engineering Department, University of Houston, Houston, TX 77004, and also with the Department of Computer Science and Engineering, Kyung Hee University, Yongin-si, Gyeonggi-do 17104, Rep. of Korea, email:{\{hanzhu22\}}@gmail.com}}

\maketitle

\begin{abstract}
Semantic communication has gained significant attention from researchers as a promising technique to replace conventional communication in the next generation of communication systems, primarily due to its ability to reduce communication costs.
However, little literature has studied its effectiveness in multi-user scenarios, particularly when there are variations in the model architectures used by users and their computing capacities. 
To address this issue, we explore a semantic communication system that caters to multiple users with different model architectures by using a multi-purpose transmitter at the base station (BS).
Specifically, the BS in the proposed framework employs semantic and channel encoders to encode the image for transmission, while the receiver utilizes its local channel and semantic decoder to reconstruct the original image.
Our joint source-channel encoder at the BS can effectively extract and compress semantic features for specific users by considering the signal-to-noise ratio (SNR) and computing capacity of the user. 
Based on the network status, the joint source-channel encoder at the BS can adaptively adjust the length of the transmitted signal.
A longer signal ensures more information for high-quality image reconstruction for the user, while a shorter signal helps avoid network congestion.
In addition, we propose a hybrid loss function for training, which enhances the perceptual details of reconstructed images.
Finally, we conduct a series of extensive evaluations and ablation studies to validate the effectiveness of the proposed system. 
\end{abstract}


% \fi
\IEEEpeerreviewmaketitle

\section{Introduction}

\IEEEPARstart{I}{n} recent years, proliferation of connected devices and their high demand for data have created significant challenges for traditional communication systems. This massive growth in device connectivity has led to a decrease in the amount of available bandwidth allocated to each user, ultimately resulting in high latency during communication processes.
In addition, the growing reliance on AI-powered devices and applications, such as autonomous vehicles, online gaming, financial trading systems, and telecommunications \cite{10004947}, also demands stable and low latency communication.
As a result, managing the network bandwidth to meet the intricate requirements of users becomes increasingly challenging as the number of users grows.
The existing communication technologies have nearly approached the Shannon physical-layer capacity limit \cite{luo2022semantic}, which is the main reason to motivate researchers to envision the potential solution for the future generation of wireless communication systems. One propitious idea is to reduce the amount of data transmission by only transmitting the important latent feature within the data. This forms the core idea behind semantic communication, which has been widely considered as a promising technique to resolve the challenges of exponential growth of network traffic, as well as new disruptive services and applications in 6G \cite{jiang2021road}.

To have a better understanding of the design and ultimate goal of semantic communication, let's start by exploring the history of classical communication, the system currently in use. Then a comprehensive discussion is provided on how semantic communication can effectively overcome the limitations of conventional communication, addressing the existing bottlenecks.
According to Shannon and Weaver \cite{shannon1949mathematical}, there are three levels of communication problems: 1) \textit{engineering problems}, 2) \textit{semantic problems}, and, 3) \textit{effective problems}. 
In the \textit{engineering problem} - communication level one, the objective is to accurately reconstruct all the symbols transmitted by the sender at the receiver side. The step-by-step process of this classical communication can be described as follow: the sender generates a message, which is then compressed and encoded by a channel encoder before being modulated to an analog signal and transmitted to the receiver. The purpose of the channel coding stage is to protect the data from being corrupted in the transmission channel. At the receiver site, the channel decoding process removes the noise from the received signal and recovers the original signal. It is apparent that the evaluation metric for this level of communication involves measuring the disparity between the transmitted sequence and the received sequence at the bit level.

The second level of the communication problem, known as the \textit{semantic problem}, pertains to the receiver's ability to accurately interpret the received message compared to its original meaning conveyed by the transmitter.
To provide a concise term for semantic problems in wireless communication, researchers commonly refer to it as semantic communication.
The ultimate objective of semantic communication is for the receiver to understand the main content or the meaningful information of the message without the need to transmit full-size data.
In order to achieve this, both the transmitter and receiver must have the ability to understand the meaning of the source data.
Specifically, the transmitter needs to be aware of the meaning of data in order to extract semantic information and eliminate the redundant or unnecessary information.
By transmitting only the semantic information, the amount of data transmission through the wireless channel can be significantly reduced, thereby decreasing the transmission time for serving each user \cite{dang2023semantic}.
At the receiver site, they receive only the shortened version of the message, and their task is to interpret it using their knowledge to fully understand the semantic meaning of the original message. 
Therefore, the efficiency of semantic communication cannot be evaluated by the number of bit differences between the transmitted signal and received signal but by the differences in meanings.
For instance, in the context of semantic communication for text, if the original message at the transmitter side is \textit{``I have an expensive automobile"} and the interpreted message at the receiver side is \textit{``I have a costly car"}, even though the wording differs, the main meaning of the sentence is preserved. Therefore, this system can be considered successful for semantic communication. 
On the other hand, if we were to evaluate the above system as in level one communication, this system would be deemed as a failure due to the differences in wording between the two sentences.
The third and final level of communication is the effective problem, which goes one step further compared to the second level. It not only requires the receiver to understand the semantic meaning of the message but also to perform certain tasks based on the understanding. Consequently, the proficiency of this last level of communication can be determined by the success of the tasks performed by the receiver.

In this paper, we focus on the existing problems in the second-level communication system.
Extracting the semantic meaning from source data was a challenging task a few decades earlier due to the lack of appropriate techniques. However, several solutions have been recently proposed to tackle this issue thanks to the breakthrough in deep learning models.
For speech transmission, \cite{weng2021semantic} utilized the squeeze-and-excitation (SE) network to transmit speech audio, achieving superior performance in speech signal metrics compared to conventional communication methods.
Another study \cite{han2022semantic} proposed a two-stage training scheme to speed up the training time. 
The transmission of semantic meaning in the text modality has been recognized as a more challenging task compared to the speech signal. Nevertheless, a number of works have been proposed to tackle this problem \cite{xie2021deep,peng2022robust}. Expanding beyond the text and speech modalities, researchers in \cite{kang2022personalized,zhang2023optimization,hu2023robust} have proposed various semantic communication models for transmitting images over a noisy wireless environment. However, it is worth noting that all of the aforementioned studies proposed communication systems with only one receiver, which is unrealistic in the real-world environment.
A couple of works have considered the multi-user scenario in semantic communication \cite{hu2022one,xie2021task}. In \cite{hu2022one}, the authors focused on designing a system for broadcasting to multiple users, while \cite{xie2021deep} attempted to solve the multi-modality problem. Nevertheless, none of these works considered the variations in the models deployed at the user sites.

To the best of our knowledge, this is the first study to consider the multi-user scenario with different computing capacities in a semantic communication system, particularly focusing on the data reconstruction task. Moreover, unlike previous approaches that fix the length of transmitting signals, our proposed approach allows the BS to dynamically control the signal length based on the network status.
While examining the reconstructed images produced by the network trained using the mean square error (MSE) loss, we observed that it fails to meet the criteria of the human visual system. To address this limitation, we introduce an alternative training objective tailored for this particular requirement.
The main contribution of our paper can be summarized as follows:
\begin{itemize}
 \item We propose a semantic communication system that considers one BS and multiple users. The BS uses a semantic encoder to extract meaningful information from the data and compresses it using a channel encoder before transmitting it to the users. Conversely, users perform the reverse process to reconstruct the complete data from semantic information. As our system allows different users to deploy different decoder architectures, the transmitter at the BS needs to encode the information in a versatile manner to serve all users.
 
 \item We adopt a different training objective instead of the MSE between the original image and the reconstructed one. Specifically, we propose a hybrid loss, which combines different loss functions: multi-scale structural similarity index measure (MS-SSIM), absolute error (AE), and MSE. With this objective, our system is able to reconstruct images with improved perceptual details compared to those obtained using the MSE loss while also obtaining better scores in various evaluation metrics. 
 
 \item At the BS, we introduce a semantic encoder with the associated target embedding, which guides the encoder to extract the semantic features specialized for one particular decoder. In addition, our channel encoder leverages both the SNR information and the network communication status to encode the semantic feature vector and adjust its length.

\item Finally, we conduct a series of simulations to evaluate the efficiency of our proposed method, compare it against the benchmark scheme, and show the improvement in the visual perspective of the reconstructed images.
\end{itemize}

The rest of this paper is structured as follows. We provide a brief overview of related works in Section \ref{Related}. Following that, Section \ref{System} provides a more detailed explanation of our system model. In Section \ref{Proposed}, we delve into our proposed solution, providing an in-depth demonstration. Then, the simulation results are presented in Section \ref{Performance}, and based on the findings, we draw a conclusion in Section \ref{Conclusion}.

\section{Related works}\label{Related}

\subsection{Semantic Communication}
In this section, we provide an overview of the literature related to semantic communication. The authors in \cite{bourtsoulatze2019deep} recognized the capability of deep neural networks for image compression and channel noise estimation, and proposed a deep auto-encoder model to transmit images over a wireless channel. Building upon this, \cite{xie2021deep} proposed a transformer-based joint semantic-channel coding system called DeepSC. DeepSC, originally designed for text transmission, aimed to maximize the system capacity while preserving the meaning of sentences. In \cite{weng2021semantic}, the DeepSC was further improved by incorporating a squeeze-excitation network to transmit speech signals. Their proposed model outperformed the conventional communication system regarding speech signal metrics and speech distortion. Another direction of semantic communication research focused on goal-oriented approaches. In \cite{farshbafan2022common}, the authors proposed a semantic communication system for performing a task with the shortest length of transmitted messages using curriculum learning. Attention-based reinforcement learning approach was used in \cite{wang2022performance} to select the most informative triplet for transmission due to the limited resource block from the BS to each user. In \cite{thomas2022neuro}, the authors considered causal reasoning capabilities of the speaker and listener, which could be explained as the ability to comprehend and employ logical reasoning when analyzing data, including its generation process. They adopted generative flow networks (GFlowNet) to achieve the causal reasoning for emergent semantic communications with fewer data. 

\subsection{Image Transmission in Semantic Communication}
Semantic communication systems have been extensively studied in the context of image transmission. However, the presence of stochastic noise in the wireless channel within real environments posed a significant challenge that could lead to performance degradation in these systems. Recognizing this issue, \cite{xu2021wireless} proposed Attention DL-based joint source channel coding (ADJSCC), which is more tolerant to noise in the real environment. It could effectively operate under different SNR levels by adjusting the number of bits allocated to the channel encoder and source encoder. DeepJSCC with feedback was introduced in \cite{kurka2020deepjscc}, where the received signal was fed back to the transmitter, allowing the system to obtain information about the channel noise and subsequently eliminate its effects. In contrast to previous works, authors in \cite{hu2023robust} considered not only the channel noise but also the semantic noise in the system by using adversarial training. They proposed a masked vector quantized-variational autoencoder to enhance the system's robustness against both types of noise. Unlike the previous approaches that optimized the network using the MSE loss, we propose an alternative loss function that can produce more visually pleasing images. 

\subsection{Transformer in Semantic Communication}

Deep learning models have shown remarkable performance in discovering complex patterns within high-dimensional data like text and images \cite{lecun2015deep}. Recently, the transformer architecture \cite{vaswani2017attention} has been recognized as a breakthrough in tackling natural language processing tasks. Leveraging the attention mechanism, transformers excel at capturing long-range dependency better than recurrent neural networks (RNN). With the creation of the vision transformer (ViT)\cite{dosovitskiy2020image}, they have also found success in various vision tasks. Observing these developments, \cite{xie2022task} utilized transformers for the encoder and decoder in semantic communication. Furthermore, \cite{yang2023witt} proposed an end-to-end semantic communication system using the shifted window (Swin) transformer for the deep joint semantic channel encoder and decoder. However, it is worth noting that \cite{yang2023witt} only considered a single-user scenario. While \cite{xie2022task} addressed multiple users but did not consider the difference in computing capacity. In contrast, our work consider the multi-user scenario with different computing capacities while also allowing the dynamic changes in the compression ratio based on the network traffic.

\begin{figure}[t!]
    \centering
    \includegraphics[width=0.9\linewidth]{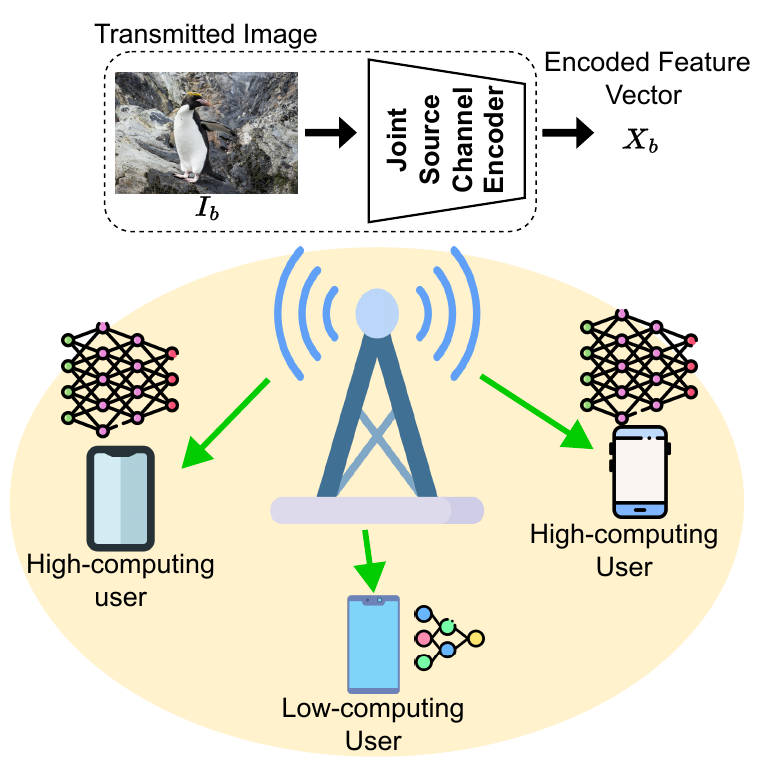}
    \caption{System model of semantic communication.}
    \label{SystemModel}
\end{figure}

\section{System model}\label{System}

As shown in Fig.~\ref{SystemModel}, we consider a semantic communication system, which consists of a BS and a set $\mathcal{K}$ of $K$ users. Similar to prior semantic communication systems, our main objective is to reduce the number of transmitted bits to the users by leveraging the semantic meaning within the data. At the same time, it is crucial to ensure that users can maintain satisfactory performance for their tasks even with the shortened signal length.
In our approach, we consider the image recovery task at the users' sites using their local decoder models. In a conventional communication system, the transmitter pre-processes and formats the image before compressing it into a codeword sequence without considering its semantic information. In contrast, semantic communication extracts the main information from the image before passing it through the compression and the channel coding stages for transmission. Additionally, we take one step further by considering the receiver device's capacity and channel condition in our approach. These aspects are discussed in more detail in the following sub-sections. 
\begin{figure*}[t!]
    \centering
    \includegraphics[width=0.98\textwidth]{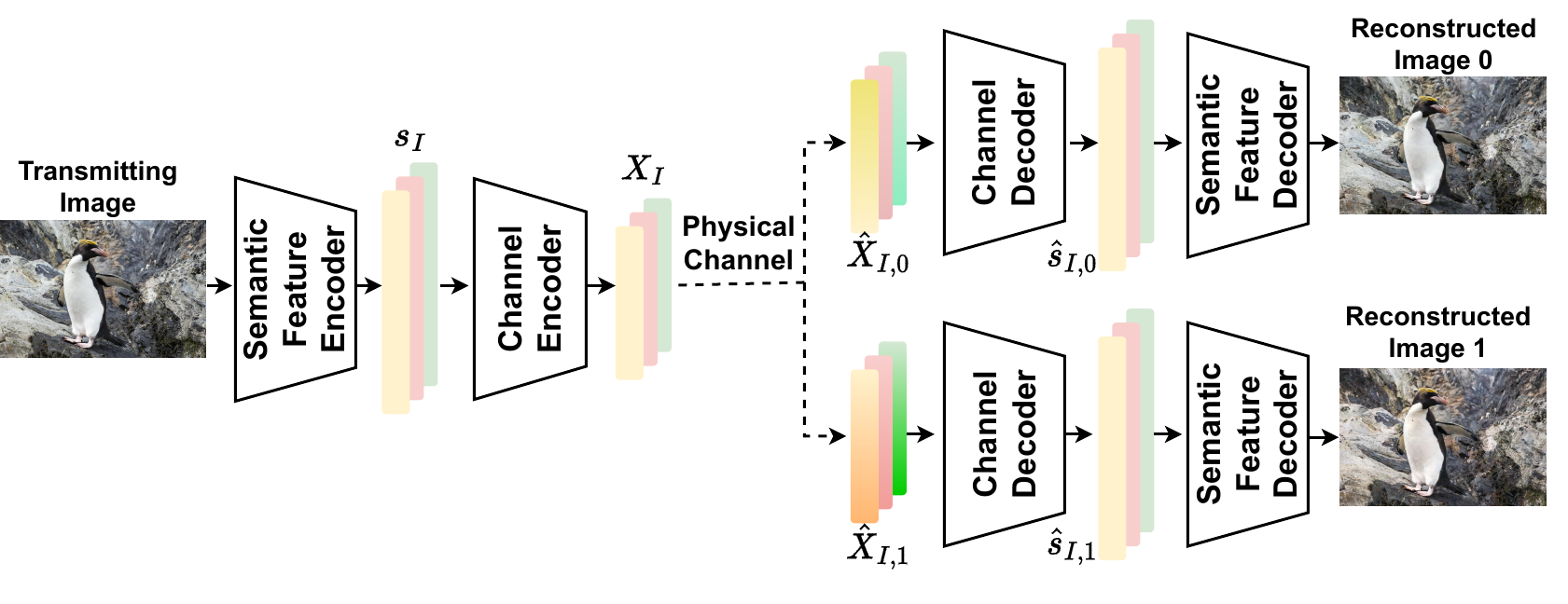}
    \caption{Image reconstruction task for muti-users with different computing capacity in semantic communication scenario.}
    \label{img1}
\end{figure*}

\begin{table}[t!]
	%	\caption{Summary of Key Notations}	
	\caption{Summary of Notations}
	\renewcommand\arraystretch{1.3}
	\begin{center}
		\begin{tabular}{|m{1.5cm}|m{6cm}|}
			%		\begin{tabular}{|p{1.5cm}|p{6.5cm}||p{2cm}|p{6.5cm}|}
			\hline
			\hfil \textbf{Notation} & \hfil \textbf{Definition}\\ \hline \hline

   			\hfil $I$ & The transmitting Image \\ \hline
   			\hfil $s_{I}$ & The extracted semantic features vectors from $I$ \\ \hline
   			\hfil $X_{I}$ & The encoded symbol of semantic features\\ \hline
         	\hfil $g_{k}$ & The combination information of CBR and SNR\\ \hline
                \hfil $E_{\alpha}$ & The semantic encoder and its parameters $\alpha$ at the BS\\ \hline
                \hfil $C_{\beta}$ & The channel encoder and its parameter $\beta$\\ \hline
                \hfil $\hat{X}_{I,k}$  & The received encoded symbol for image $I$ at user $k$\\ \hline
                \hfil $\hat{s}_{I,K}$  & The estimated semantic feature vectors for image $I$\\\hline
                \hfil $\hat{I}_{k}$ & The reconstructed image at user $k$ \\\hline
                \hfil $C^{-1}_{\phi_{k}}$ & The channel decoder at user $k$ and its parameters $\phi$\\ \hline
                \hfil $D_{\theta_{k}}$ & The semantic decoder at user $k$ and its parameters $\theta$\\ \hline
                \hfil $\mathbb{SNR}$ & The SNR set for training process\\\hline
                \hfil $\mathbb{CBR}$ & The CBR set for training process\\\hline
		\end{tabular}
		\label{tab1}
	\end{center}
\end{table}

\subsection{The joint semantic-channel encoder at the BS}

The BS in a semantic communication system typically consists of two modules: the semantic feature encoder and the channel encoder. The semantic feature encoder, which also can be referred to as the source encoder, takes an image as its input and outputs a set of important features extracted from the image. We denote the input image as $I \in \mathbb{R}^{n}$, where $n$ is the length of the vector and can be calculated by image height $\times$ image weight $\times$ image channel. The feature vectors extracted from $I$ can be denoted as:
\begin{equation}
   \boldsymbol{s}_{I}= E_{\alpha}(I),
   \label{eq1}
\end{equation}
where $E_{\alpha}(\cdot)$ is the semantic encoder network with the parameter set $\alpha$. With the extracted feature vectors, we feed them into a channel encoder to protect the transmitted symbol from channel noise and interference in the wireless environment. The parameter set of the channel encoder will be denoted as $\beta$, while the encoded symbols are being presented as: 
\begin{equation}
    X_{I}= C_{\beta}(\boldsymbol{s}_{I}) \in \mathbb{R}^{k},
    \label{eq2}
\end{equation}
where $C_{\beta}(\cdot)$ is the channel encoder with the parameter $\beta$, and $k$ denotes the length of $X_{I}$. While enhancing the signal tolerance to interference, the channel encoder also facilitates the compression of semantic features to a lower dimension. We calculate the compression ratio by dividing the transmitting size by the size of the image, which is $k/n$. 
Although it is possible to compress the semantic feature vector further for shorter transmission length, there exists a trade-off between the compression size and the quality of the reconstructed image.
Most existing works only consider one transmitter and one receiver, and therefore disregarding the receiver's computing capacity during the encoding process. In the case of multiple receivers, each user device may have different computing capacities and energy to decode the received signal from the transmitter. For example, a user with high-computing capacity may strive to reconstruct a high-quality image, whereas a user with limited computing capacity may aim to generate a coarse representation of the image without intricate details. To this end, we propose a conditional semantic encoder, which considers the receiver's resource capacity as extra information during the feature extraction process. Therefore, Eqn. \eqref{eq1} can be rewritten as follows:
\begin{equation}
   \boldsymbol{s}_{I,k}= E_{\alpha}(I|c_{k}),
   \label{eq3}
\end{equation}
where $c_{k}$ is the computing information of receiver $k$, and $s_{I,k}$ is the extracted feature vector for receiver $k$. Consequently, the encoded feature vectors for a specific user will be denoted as follows:
\begin{equation}
    X_{I,k}= C_{\beta}(\boldsymbol{s}_{I,k}|g_{k}),
    \label{eq4}
\end{equation}
where $g_{k}$ contains information about the channel from BS to receiver $k$, particularly the signal-to-noise ratio (SNR) and also the network demand. The feedback information about the SNR to the transmitter gives the channel encoder knowledge about the possible noise amplitude in the environment. While the network demand information is helpful in adjusting the CBR to improve the reconstructed image quality (high CBR) or ease the network traffic (low CBR). 

\subsection{Joint semantic-channel encoder at receivers}
At the receiver site, users are equipped with the channel decoder and the semantic decoder. It is important to note that the encoded feature vectors are transmitted over a wireless environment, in which channel distortion and noise are inevitable. Firstly, we study the additive white Gaussian noise (AWGN) channel, in which the received vectors at the user $k$ can be denoted as: 
\begin{equation}
    \hat{X}_{I,k}=Y_{I,k}= X_{I,k} + N_{k},
    \label{AWGNequation}
\end{equation}
where $Y_{I,k}$ is the received signal at user $k$ for the transmitted image $I$, and $N_{k}$ is the complex noise vector, whose elements follow the $\mathcal{CN}(0,\sigma^{2}\boldsymbol{I})$ distribution. When the Rayleigh fading channel is considered, the received signal is given by the following equation:
\begin{equation}
    Y_{I,k}= X_{I,k}H_{k}+ N_{k},
\end{equation}
where $H_{k}$ is the fading coefficient between the transmitter and receiver $k$. With the knowledge of channel estimation and zero forcing detector, the signal can be transformed back to \eqref{AWGNequation} as follows \cite{xie2021task}:
\begin{equation}
    \hat{X}_{I,k}=(H^{H}H)^{-1}H^{H}Y^{I,k}= X_{I,k} + \hat{N}_{k},
    \label{fadingtowhite}
\end{equation}
where $\hat{X}$ is the estimated feature vector at receiver $k$, while $\hat{N}_{k}$ is the noise in Rayleigh fading channel. Using this technique, we can convert the channel effect from a multiplicative operation to an additive operation, thereby significantly reducing the learning burden. Receiver $k$ uses its channel decoder to recover estimated semantic feature vectors from the estimated encoded symbol $\hat{X}_{I,k}$ as follows:
\begin{equation}
    \hat{s}_{I,k}= C^{-1}_{\phi_{k}}({\hat{X}_{I,k}}),
\end{equation}
where $\hat{s}_{I,k}$ is the estimated feature vector of the image $I$ and $C^{-1}_{\phi}(\cdot)$ is the channel decoder of the user for decoding the received message. The main objective of the proposed system is to reconstruct the original image at the receiver site using the feature vector from the channel decoder. To achieve this, the receiver employs a semantic decoder to transform the semantic features vector into the estimated image, which is denoted as: 
\begin{equation}
    \hat{I}_{k}= D_{\theta_k}(\hat{s}_{I,k}),
\end{equation}
where $D_{\theta_{k}}(\cdot)$ denotes the semantic decoder of user $k$ with its parameters $\theta_{k}$. The difference between the original image and the one reconstructed by user $k$ can be measured using MSE loss as shown below \cite{xu2021wireless}.
\begin{equation}
    d(I, \hat{I}_{k})= \frac{1}{n}\sum_{i=1}^{n}(I_{i}-\hat{I}_{i})^2,
\end{equation}
where $n$ is the size of the image, which is a product of height, width, and the number of channels in the image. As previously mentioned, different users may have different computing capacities, which can influence the complexity of their neural network models such as the number of layers. Most prior works focused only on developing the semantic encoder and decoder while using simple and small neural networks for channel encoding and decoding. Therefore, it can be assumed that a significant portion of computing capacity is dedicated to the semantic decoding process. With this assumption, we model the semantic decoder of users with a high computing capacity to be equipped with a more powerful model compared to those with low-computing capacity. The quality of the reconstructed image at the receiver depends on many factors, including channel noise, the extracted feature vector from the semantic encoder, and the computing capacity itself.

In Fig.~\ref{img1}, we illustrate the general design of a multi-user semantic communication system, which contains a single transmitter and multiple receivers. Specifically, we consider two users (i.e., $k=[0,1]$).
For example, if the BS targets the user $0$, the image undergoes encoding by two consecutive encoders to get feature vector $X_{I,0}$. Then it is transmitted over the physical wireless channel and is affected by the noise from the environment. The user $0$ receives $\hat{X}_{I,0}$ and reconstructs the image using two consecutive decoders.

\begin{figure*}[!t]
    \centering
    \includegraphics[width=1\textwidth]{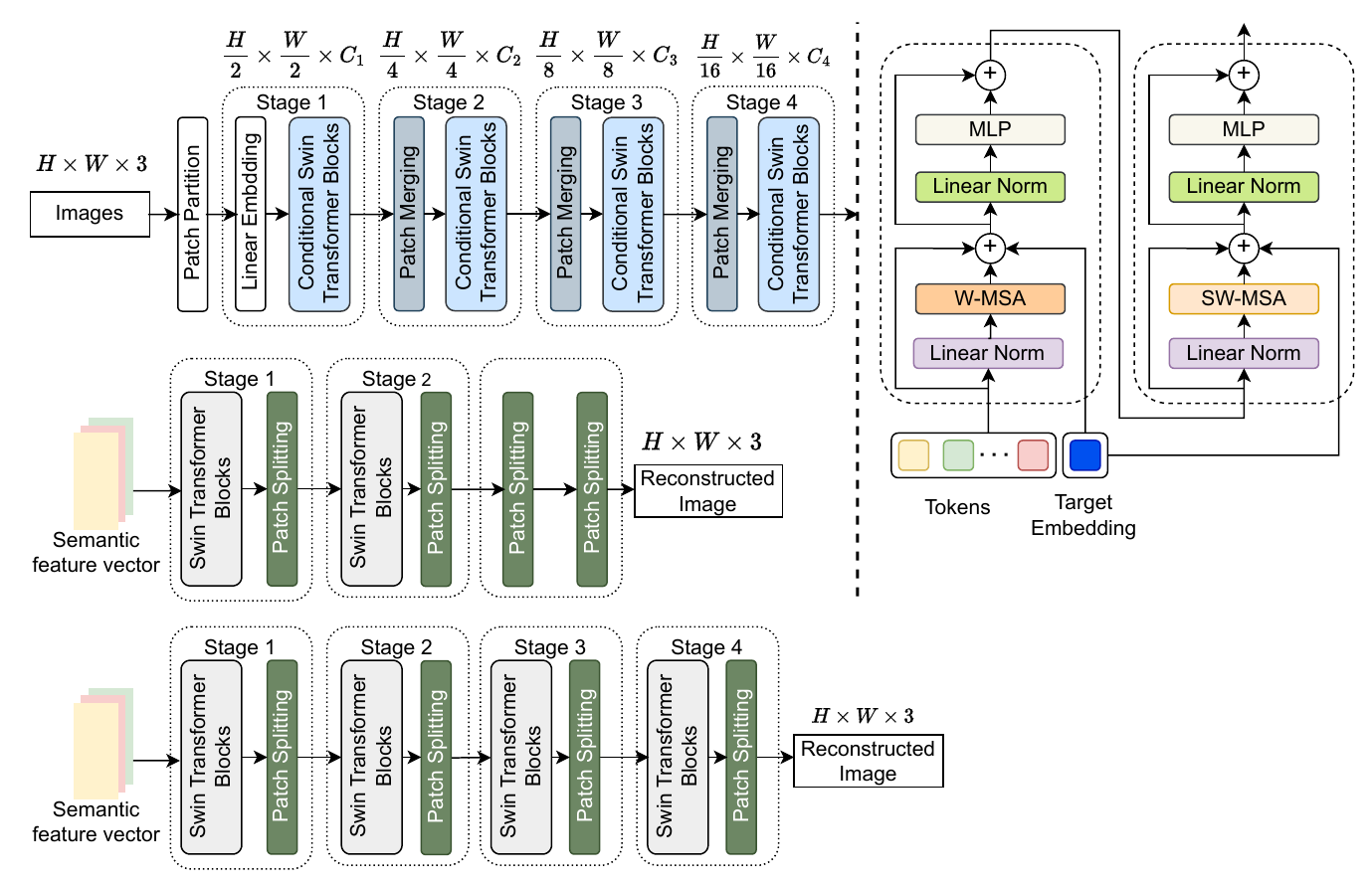}
    \caption{The architectures of the semantic encoder, low-computing semantic decoder, and high-computing semantic decoders are presented on the left-hand side. On the right-hand side, two successive Swin Transformer blocks are modified to inject decoder information for the encoder, while the structure remains the same for the blocks in both decoders.}
    \label{Swin}
\end{figure*}
\section{proposed solution}\label{Proposed}

A multi-user semantic communication system as shown in Fig.~\ref{img1} may face several challenges. Firstly, the encoder at the BS needs to encode data in a way that can be utilized by multiple users who are equipped with different decoder models. Secondly, the BS has to consider the dynamic channel conditions of the wireless environment and the status of the network traffic in the transmission process. For each challenge addressed above, we propose a novelty approach to tackle it and finally propose a new loss function for the training process. 
In Fig.~\ref{Swin}, we illustrate the proposed architectures of the semantic encoder and two semantic decoders with different computing capacities. These models are constructed based on the state-of-the-art vision transformer - Swin Transformer\cite{liu2021swin}. There are a couple of advancements in the Swin Transformer compared to the ViT\cite{dosovitskiy2020image}, so we chose this model. First of all, the computation complexity of the Swin Transformer is linear with the image size due to self-attention in the local window, it is quadratic for the conventional ViT model\cite{zhu2021long}. Moreover, Swin Transformer has the ability to produce a hierarchical representation by the patch merging. For the semantic encoder, we make changes to the patch size and the dimension of each patch after the merging operation. As shown in the right-hand side of Fig.~\ref{Swin}, we specifically incorporate the target embedding vector into the self-attention block. To be precise, it is injected after the W-MSA or SW-MSA blocks. The primary role of the target embedding vector is give information to the semantic encoder about the specific target decoder. The design is motivated by the assumption that the encoder can extract the semantic features that are particularly suitable for a specific semantic decoder when it has knowledge of the user's computation capacity (high-computing or low-computing model). This design also incorporate some degree of privacy preservation by preventing untargeted decoders from reconstructing high-quality images in case they receive unwanted signals. For the semantic decoder, we follow a reverse architecture of the encoder, starting by inputting the semantic feature vector into Swin Transformer blocks and then splitting it into patches. The high-computing semantic decoder comprises four stages of transformer architecture, while the low-computing one only contains two stages, as shown in Fig.~\ref{Swin}. Besides, the high computing decoder has more transformer blocks in each stage than the other. More details on the design of Swin Transformer model can be found in \cite{liu2021swin}.

\subsection{Training process for multi-user having two different computing capacities}

During the training process, we first alternatively couple the encoder with one decoder and train in an end-to-end manner. To differentiate the training process of two decoders, we associate the low-computing decoder with the target embedding vector value of zero, and the high-computing decoder with value one. Losses from the low-computing and high-computing decoders can be calculated as $\mathcal{L}_{0}=E(I-\hat{I}_{0})$ and $\mathcal{L}_{1}=E(I-\hat{I}_{1})$, respectively. When training the encoder and low-computing decoder, we update the parameters $\alpha$, $\beta$, $\phi_{0},$ and $\theta_{0}$ using the following loss value: 
\begin{equation}\label{updateloss1}
    \mathcal{L}= \mathcal{L}_{0}.
\end{equation}
After that the common encoder and the high-computing decoder are updated based on the following loss:
\begin{equation}\label{updateloss2}
    \mathcal{L}= \mathcal{L}_{1}.
\end{equation}
The above loss functions are used to target one particular decoder at a time. If we want the semantic encoder to produce the semantic features that can be used for both decoders (i.e., the broadcast scenario), then the following loss is proposed for the parameter update:
\begin{equation}\label{updateloss3}
    \mathcal{L}= (\mathcal{L}_{0} + \mathcal{L}_{1})/2.
\end{equation}
In this case, we update parameters for both decoders and encoder simultaneously in one training epoch. Therefore, we do not inject the target embedding vector to the semantic encoder, in this instance. 
\begin{figure*}[!t]
    \centering
    \includegraphics[width=1\linewidth]{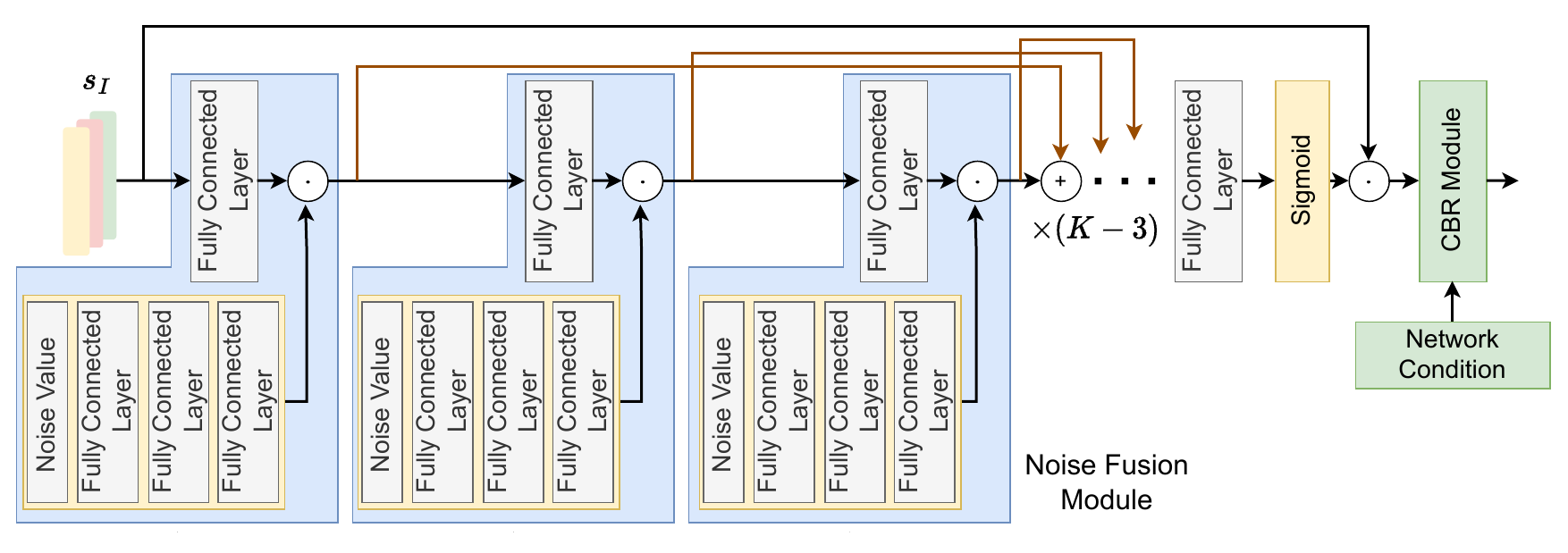}
    \caption{Channel Encoder with K noise fusion modules and the CBR module to adjust the compression ratio.}
    \label{ChannelEncoder}
\end{figure*}

\subsection{The hybrid loss function}
Most existing works related to the deep joint semantic and channel coding for semantic image transmission consider the MSE loss as their objective for the training process. MSE is a widely used loss function for neural network training due to its continuously-differentiable properties across any domain. In addition, one of the metric to evaluate the image transmission performance of semantic communication is the peak signal-to-noise ratio (PSNR), which is inversely proportional to the MSE.
\begin{equation}
    \textrm{PSNR}= 10\log_{10}\frac{\textrm{MAX}^{2}}{\textrm{MSE}},
    \label{PSNRreverseMSE}
\end{equation}
where $\textrm{MAX}$ denotes the maximum possible value of the image pixels \cite{bourtsoulatze2019deep}. For an image with 8-bit pixels in a single color channel, the $\textrm{MAX}$ value is 255. 
Therefore, minimizing the MSE leads to maximizing the PSNR. Through a series of simulations, we have observed that, in certain cases, a high PSNR value for the reconstructed image does not necessarily ensure the preservation of visual details. To address this limitation and maintain the structural integrity of the reconstructed image, we proposed a hybrid loss: 
\begin{equation}
    \label{eq:hybrid_loss}
    \mathcal{L}^{\textrm{Hyb}}= \gamma\cdot \mathcal{L}^{\textrm{MS-SSIM}} + (1-\gamma)\cdot\mathcal{L}^{\ell_{1}}+ \epsilon\cdot\mathcal{L}^{\ell_{2}},
\end{equation}
where $\mathcal{L}^{\textrm{MS-SSIM}}$ denotes the multi-scale structural similarity index measure (MS-SSIM) loss, while  $\mathcal{L}^{\ell_{1}}$ and $\mathcal{L}^{\ell_{2}}$ denote absolute error (AE) loss, and MSE loss, respectively. Here, $\gamma$ and $\epsilon$ are coefficients of the losses to prevent any loss from dominating the objective. $\mathcal{L}^{\textrm{Hyb}}$ will be used as $\mathcal{L}_{0}$ and $\mathcal{L}_{1}$ in \eqref{updateloss1}, \eqref{updateloss2} and \eqref{updateloss3}. We discuss each component of $\mathcal{L}^{\textrm{Hyb}}$ below.

\subsubsection{Structural Error - SSIM and MS-SSIM losses}
SSIM has been proposed as a metric for measuring the similarity between two images, where the original idea behind the technique is based on spatially close pixels having strong inter-dependencies. These dependencies contain important information about the structure of the objects in the visual scene \cite{wang2005structural}. The detailed algorithm for calculating the SSIM value, which involves a sliding window, can be found in \cite{wang2004image}.
\begin{equation}
    \textrm{SSIM}(x,y)= \frac{(2\mu_{x}\mu_{y}+c_{1})(\sigma_{xy}+c_{2})}{(\mu^{2}_{x}+\mu^{2}_{y}+c_{1})(\sigma^{2}_{x}+\sigma^{2}_{y}+c_{2})},
\end{equation}
where $\mu_{x}$ and $\mu_{y}$ denote the pixel mean of window $x$ from original image and window $y$ from reconstructed one, respective, while $\sigma^{2}_{x}$ and $\sigma^{2}_{y}$ are the corresponding variances. The covariance of two windows is $\sigma_{x,y}$. The loss function for using this metric is: 
\begin{equation}
    \mathcal{L}^{\textrm{SSIM}}(P)= \frac{1}{\lvert P\lvert} \sum_{p\in P} (1- \textrm{SSIM}(x,y)),
\end{equation}
where $P$ is the total number of batches in the image. Utilization of SSIM allows us to produce visually appealing images with improved structure. However, determining an appropriate window size poses a challenge. A large window size preserves the global structure information but the local structure information may suffer. Conversely, if the window size is small, noise is more likely to appear at the edges between windows, and preserving the global structure becomes more challenging. Therefore, \cite{wang2003multiscale}  introduced multi-scale SSIM to incorporate image details at different resolutions by iteratively applying a low-pass filter and downsampling by a factor of two. Then, SSIM is calculated for the images at the original scale and the downsampled scales. Therefore, the loss function for MS-SSIM is given as:
\begin{equation}
    \mathcal{L}^{\textrm{MS-SSIM}}= 1- \prod_{j=1}^{M}\textrm{SSIM}_{j}(x,y),
\end{equation}
where $\textrm{SSIM}_{j}$ denotes the mean SSIM value at downsampled scale $j$. When $j$= 1, the image is at the original size. In this paper, we lightly explain the reasons why we chose the MS-SSIM as one of the components of our loss function. A more detailed explanation of the design and the idea behind  MS-SSIM can be found in \cite{wang2003multiscale}.
\subsubsection{Mean Square Error - MSE}
The advantages of MSE have been proved in previous works and mentioned in the preceding section. Our objective is not only to ensure that the reconstructed image possesses the same structure as the original image but also to maintain consistency in pixel values between the two images. Therefore, we incorporate MSE as a component in our loss function. 

\subsubsection{Absolute Error - AE}
While the MSE is highly sensitive to outliers points in an image and can be unstable when dealing with them, AE is known to be more robust and less not affected by outliers. Moreover, authors in \cite{zhao2016loss} found that combining AE loss and $\textrm{MS-SSIM}$ yields promising results in the field of image restoration.

\subsection{Dynamic Compression Ratio}
So far, we have discussed the semantic encoder and decoders, responsible for extracting the semantic meaning of the image and reconstructing it. Meanwhile, the channel encoder is in charge of encoding the extracted semantic feature vector for transmission. The encoding process plays a crucial role in protecting the transmitted signal against noise and interference in the wireless environment. We leverage the SNR information to perform the channel encoding process effectively as in \cite{yang2023witt,xu2021wireless}. Moreover, we propose a channel encoder that is capable of outputting different sizes of signal to adapt to the network congestion status. For example, when there are many images to be transmitted in the queue, we can compress the image a little bit more to deliver it faster. On the other hand, if the BS is not in high demand for fast transmission, we can compress the image slightly less aggressively to achieve high-quality reconstructed images. The reason for this design is that the compression bit ratio directly affects the quality of the reconstructed image at the receiver site. Although the channel encoder can use a low compression ratio to aggressively shorten the transmitted signal length, the channel decoder may face challenges in decompressing and removing the channel noise with the shortened data signal. The channel decoder may have a better chance at noise removal and decompression with a longer signal.

The architecture of the proposed channel encoder is shown in Fig.~\ref{ChannelEncoder}. It is composed of $K$ noise fusion modules and one compression bit ratio (CBR) module.
The noise fusion module associates each semantic feature with a weight and computes the dot product between them and output from weighted noise values. Using the SNR information, we can derive a set of potential noise values, which are passed through three interconnected layers. The resulting outputs are referred to as weighted noise values. In addition, we apply a two-block skip operation to the noise fusion module's output, denoted by the brown line in the figure.
\begin{table*}[t]
\centering
\caption{the ms-ssim performance of two decoders in different scenarios under various channel conditions.}
\renewcommand{\arraystretch}{1.1}
\label{Targetandbroadcast}
\begin{tabular}{|c|ccc|ccc|}
\hline
          & \multicolumn{3}{c|}{Low-computing Decoder}                                                                                  & \multicolumn{3}{c|}{High-computing Decoder}                                                                                  \\ \hline
SNR values & \multicolumn{1}{c|}{Non-targeted}       & \multicolumn{1}{c|}{Broadcast}           & Targered            & \multicolumn{1}{c|}{Non-targeted}       & \multicolumn{1}{c|}{Broadcast}           & Targered            \\ \hline
0 dB      & \multicolumn{1}{c|}{0.8216 $\pm$ 0.0006} & \multicolumn{1}{c|}{0.8792 $\pm$ 0.0003} & 0.8846 $\pm$ 0.0004 & \multicolumn{1}{c|}{0.8998 $\pm$ 0.0001} & \multicolumn{1}{c|}{0.9081 $\pm$ 0.0001} & 0.9135 $\pm$ 0.0002 \\ \hline
2 dB      & \multicolumn{1}{c|}{0.8713 $\pm$ 0.0009} & \multicolumn{1}{c|}{0.9154 $\pm$ 0.0001} & 0.9189 $\pm$ 0.0003 & \multicolumn{1}{c|}{0.9212 $\pm$ 0.0001} & \multicolumn{1}{c|}{0.9338 $\pm$ 0.0001} & 0.9384 $\pm$ 0.0001 \\ \hline
4 dB      & \multicolumn{1}{c|}{0.8960 $\pm$ 0.0007} & \multicolumn{1}{c|}{0.9375 $\pm$ 0.0001} & 0.9397 $\pm$ 0.0001 & \multicolumn{1}{c|}{0.9360 $\pm$ 0.0001} & \multicolumn{1}{c|}{0.9498}              & 0.9534             \\ \hline
6 dB      & \multicolumn{1}{c|}{0.9148 $\pm$ 0.0003} & \multicolumn{1}{c|}{0.9518}              & 0.9530              & \multicolumn{1}{c|}{0.9455 $\pm$ 0.0001} & \multicolumn{1}{c|}{0.9605}              & 0.9630              \\ \hline
8 dB      & \multicolumn{1}{c|}{0.9286 $\pm$ 0.0002} & \multicolumn{1}{c|}{0.9607}              & 0.9609              & \multicolumn{1}{c|}{0.9519}              & \multicolumn{1}{c|}{0.9676}              & 0.9691           \\ \hline
\end{tabular}
\end{table*}
We assume the communication network has three states: low-demand state, normal-demand state, and finally, high-demand state. This network information will be the input of the CBR module. The CBR module comprises three FC layers whose output sizes correspond to the following CBR [3/64; 4/64; 5/64], respectively. Depending on the network status, the module will choose one of three FC layers. For the channel decoder, we inverse the above process by first putting the received signal into the CBR module. Then, we move the sigmoid activation dot product to the other side of the network to output the $\hat{s}_{I,k}$. 

The details of the proposed framework can be found in Algorithm~\ref{alg:cap}. It is important to note that the algorithm presented here only illustrates a scenario with two receivers. However, the scalability of the framework to accommodate a larger number of users can be easily  resolved by associating each user with a value of $c_{k}$ and sequentially training all decoders with the multi-purpose transmitter.

% $E_{\alpha}(I|c_{k})$
\newfloat{algorithm}{t}{lop}
\begin{algorithm}
\caption{Training framework for multi-user with two computing capacities}\label{alg:cap}
\textbf{Input:} Total training epoch $M$, the transmitting image $I$, value sets of CBR and SNR.\\
\textbf{Output:} The optimal network $ E_{\alpha}, C_{\beta}, C^{-1}_{\phi_{0}},D_{\theta_{0}}, C^{-1}_{\phi_{1}} , D_{\theta_{1}}$.
\begin{algorithmic}[1]
\State Initialize the network parameters
% \newline Calculate the latency of each agent decision and determine the optimal decision X*, $t \gets 0$.
\While{$m \leq M $}
    \If{$m$ $mod$ $2$ $= 0$}
        \State $c_{k}$ = 0; freeze $C^{-1}_{\phi_{1}}$ and $D_{\theta_{1}}$; $\mathcal{L}$ $\gets$ Eq.~(\ref{updateloss1})
    \Else
        \State $c_{k}$ = 1; freeze $C^{-1}_{\phi_{0}}$ and $D_{\theta_{0}}$; $\mathcal{L}$ $\gets$ Eq.~(\ref{updateloss2})
    \EndIf
    \For{batch $\leq$ B}
        \State $g_{k}$ = $(\textrm{SNR}, \textrm{CBR}) \in $ 
 $\mathbb{SNR}, \mathbb{CBR}$.
        \State $s_{I,k}$ = $E_{\alpha}(I|c_{k})$; $X_{I,k}= C_{\beta}(\boldsymbol{s}_{I,k}| g_{k} ) $
        \State $\hat{s}_{I,0}= C^{-1}_{\phi_{0}}({\hat{X}_{I,k}})$; $\hat{I}_{0}= D_{\theta_{0}}(\hat{s}_{I,k})$
        \State $\hat{s}_{I,1}= C^{-1}_{\phi_{1}}({\hat{X}_{I,k}})$; $\hat{I}_{1}= D_{\theta_{1}}(\hat{s}_{I,1})$
        \State $\mathcal{L}_{0}$ = $\mathcal{L}^{\textrm{Hyb}}(I,\hat{I}_{0})$; $\mathcal{L}_{1}$ = $\mathcal{L}^{\textrm{Hyb}}(I,\hat{I}_{1})$ 
        \State Calculate $\mathcal{L}$ and update network parameters.
    \EndFor
    \State $m \gets m +1$. 
\EndWhile
\State Return ($ E_{\alpha}, C_{\beta}, C^{-1}_{\phi_{0}},D_{\theta_{0}}, C^{-1}_{\phi_{1}} , D_{\theta_{1}}$.).
\end{algorithmic}
\end{algorithm}
\section{Performance Evaluation}\label{Performance}

In this section, we conduct a series of experiments to analyze the effectiveness of our proposed system in different physical environment conditions while also comparing it with other benchmark schemes. We consider the AWGN with a wide range of SNR values. The Rayleigh fading channels can also be converted back to AWGN as shown in \eqref{fadingtowhite}. To the best of our knowledge, our proposal stands as a pioneering and practical application of semantic communication within the existing literature, since we consider the multi-user scenario with varying computing capabilities. Therefore, we would like to emphasize that currently there are no direct comparison schemes available for the outcomes presented in this study. We rebuilt the proposed model in \cite{xu2021wireless}, which is ADJSCC, to show the efficiency of our hybrid loss in the context of semantic communication with multi-users equipped with different DL models. 

\subsection{Training \& Evaluating dataset}\label{traningdetail}

We implement the proposed models and other benchmark schemes using Pytorch \cite{paszke2019pytorch}. In order to recognize the visual differences among reconstructed images, it is essential to use a high-resolution dataset. The DIV2K dataset \cite{agustsson2017ntire} is a suitable choice for this purpose. We use the training portion for training the network, while all the images from the testing portion are used to evaluate the model performance. On top of that, we also validate the performance with another dataset, which is Kodak\cite{franzen1999kodak}, to ensure the generalization capacity of our model. During the training process, our models are trained under a wide range of SNR vales, i.e, [1, 3, 5, 7]. The semantic encoder is comprised of four stages, each with a specific number of transformer blocks: (2, 2, 6, 2). In contrast, the high-computing decoder is composed of transformer blocks arranged as (2, 6, 1, 1). Conversely, the low-computing decoder consists of two stages, with each stage having a single transformer block. The channel encoder and decoders have seven noise fusion modules and use the Adam optimizer\cite{KingBa15} with a 0.00005 learning rate.
Due to the dynamic of the noise from the environment, we run the simulations three times and calculate the mean and standard deviation for each experiment.
\begin{figure}[t]
    \centering
    \includegraphics[width=1\linewidth]{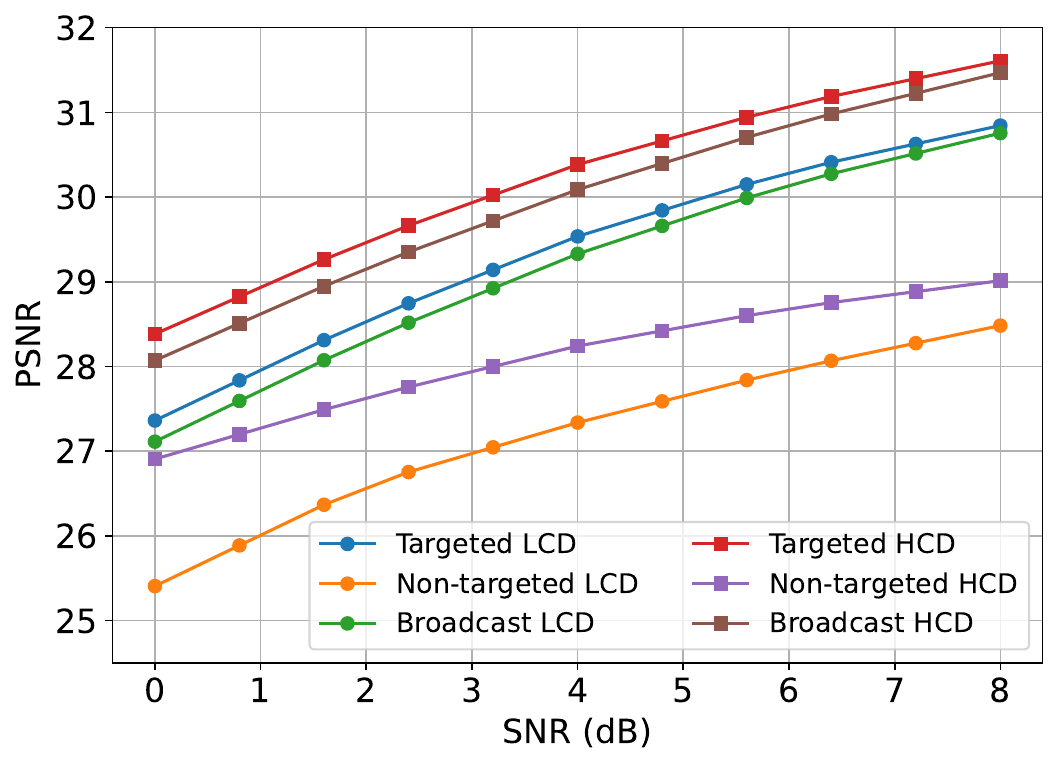}
    \caption{The PSNR results of high-computing and low-computing decoder in three scenarios: 1) the targeted decoder, 2) non-targeted decoder 3) the broadcasting case.}
    \label{PSNRtargetandbroadcast}
\end{figure}
\begin{table*}[t]
\centering
\caption{the psnr performance of decoders with mse and hybrid losses.}
\renewcommand{\arraystretch}{1.1}
\label{MSEHybridtable}
\begin{tabular}{|c|cc|cc|}
\hline
          & \multicolumn{2}{c|}{LCD}                                                             & \multicolumn{2}{c|}{HCD}                                                   \\ \hline
SNR values & \multicolumn{1}{c|}{MSE}                            & Hybrid                         & \multicolumn{1}{c|}{MSE}                  & Hybrid                         \\ \hline
0 dB      & \multicolumn{1}{c|}{27.3615 $\pm$ 0.0061}           & \textbf{27.3924 $\pm$ 0.0014}  & \multicolumn{1}{c|}{28.3832 $\pm$ 0.0031} & \textbf{28.4742  $\pm$ 0.0025} \\ \hline
2 dB      & \multicolumn{1}{c|}{\textbf{28.5477 $\pm$ 0.0058}}  & 28.5406 $\pm$ 0.0023           & \multicolumn{1}{c|}{29.4828 $\pm$ 0.0038} & \textbf{29.5758  $\pm$ 0.0031} \\ \hline
4 dB      & \multicolumn{1}{c|}{\textbf{29.55355 $\pm$ 0.0065}} & 29.5014  $\pm$ 0.0006          & \multicolumn{1}{c|}{30.3833 $\pm$ 0.0025} & \textbf{30.5145  $\pm$ 0.0016} \\ \hline
6 dB      & \multicolumn{1}{c|}{\textbf{30.3029 $\pm$ 0.0037}}  & 30.2722  $\pm$ 0.0013          & \multicolumn{1}{c|}{31.0820 $\pm$ 0.0022} & \textbf{31.2567  $\pm$ 0.0005} \\ \hline
8 dB      & \multicolumn{1}{c|}{30.8440 $\pm$ 0.0008}           & \textbf{30.8606  $\pm$ 0.0014} & \multicolumn{1}{c|}{31.6085 $\pm$ 0.0013} & \textbf{31.8036 $\pm$ 0.0023}  \\ \hline
\end{tabular}
\end{table*}

\subsection{Effectiveness of target embedding vector at the encoder}

The target embedding vector is used to inform the semantic encoder to extract the feature vectors optimized for a particular decoder only, ensuring that the unintended recipient decoder cannot interpret those features or at least it cannot produce high-quality images. In contrast, the broadcast communication allows the encoded semantic features from the transmitter to be interpreted by all the receivers in the network. In this subsection, we examine both scenarios mentioned above: target and broadcast cases. As shown in Fig.~\ref{PSNRtargetandbroadcast}, we compare the performance of the high-computing decoder(HCD) and low-computing decoder(LCD) under different channel conditions using the PSNR metric. In general, the PSNR values of both decoders increase when the condition of the physical channel improves. It is obvious that the HCD outperforms the LCD in most cases, with the exception of non-targeted HCD. We can notice that there is a significant difference in the PSNR values between targeted and non-targeted decoders. These gaps become larger as the SNR value increases, due to the rapid improvement of the targeted case compared to the slower improvement of the other. In addition, the targeted HCD slightly outperforms the broadcast HCD in every case, which proves the efficiency of the target embedding vector. Figure~\ref{PSNRtargetandbroadcast} is drawn by using the mean value only while ignoring the standard deviation due to its small value.

Besides PSNR, we also report the MS-SSIM metric for all the mentioned cases in Table. \ref{Targetandbroadcast}. We can observe that the performance of DL models is stable in every case, which is proven by the standard deviation value. With an SNR value of 0dB, the low-computing decoder achieves a value up to 0.8846 when it is targeted and only around 0.8216 for the non-targeted case. This performance difference can be considered a huge gap as the maximum value of MS-SSIM is 1. In the meantime, the low-computing decoder in the broadcast mode also obtains the high MS-SSIM value 0.8792, which is 0.005 lower compared to the targeted case at the same channel condition. As the SNR increases, indicating a better channel condition, the MS-SSIM of the decoder also significantly improves, from 0.8846 at 0 dB to 0.9609 at 8 dB. The performance of the high-computing decoder follows a similar trend while acquiring a better value due to its larger model.

\subsection{Performance gain from the hybrid loss}

\begin{table*}[t]
\centering
\renewcommand{\arraystretch}{1.1}
\caption{the performance difference in psnr under harsh and inexperience  channel conditions}
\label{PSNRunderharshenvironment}
\begin{tabular}{|c|cc|cc|}
\hline
          & \multicolumn{2}{c|}{LCD}                                                   & \multicolumn{2}{c|}{HCD}                                                   \\ \hline
SNR value & \multicolumn{1}{c|}{MSE}                  & Hybrid                         & \multicolumn{1}{c|}{MSE}                  & Hybrid                         \\ \hline
-1 dB     & \multicolumn{1}{c|}{26.6033 $\pm$ 0.0076} & \textbf{26.6907 $\pm$ 0.0020}  & \multicolumn{1}{c|}{27.7031 $\pm$ 0.0042} & \textbf{27.8134  $\pm$ 0.0037} \\ \hline
-2 dB     & \multicolumn{1}{c|}{25.6142 $\pm$ 0.0087} & \textbf{25.7889 $\pm$ 0.0050}  & \multicolumn{1}{c|}{26.8353 $\pm$ 0.0056} & \textbf{26.9571$\pm$ 0.0034}   \\ \hline
-3 dB     & \multicolumn{1}{c|}{24.2937 $\pm$ 0.0092} & \textbf{24.5803  $\pm$ 0.0088} & \multicolumn{1}{c|}{25.6646 $\pm$ 0.0072} & \textbf{25.7537  $\pm$ 0.0053} \\ \hline
-4 dB     & \multicolumn{1}{c|}{22.6123$\pm$ 0.0079}  & \textbf{22.9984  $\pm$ 0.0146} & \multicolumn{1}{c|}{24.0730$\pm$ 0.0076}  & \textbf{24.0749  $\pm$ 0.0088} \\ \hline
\end{tabular}
\end{table*}

In this particular subsection, we carry out a series of simulations to validate the efficiency of the proposed hybrid loss. In Table.~\ref{MSEHybridtable}, we compare PSNR results given by both the HCD and the LCD with different training losses. According to \eqref{PSNRreverseMSE}, MSE is inversely proportional to PSNR. Therefore, it is easy to assume the models, whose target is to minimize the MSE alone, will obtain the highest PSNR value. However, as shown in the table, the performance of our hybrid loss even produces results with slight improvements in several cases. The slight improvement observed in the hybrid loss case is a very interesting point, and one possible argument is that the training process using a single objective may become stuck in a local optimum, while the combination of different objectives in the hybrid loss allows for more dynamic network behavior.

\begin{figure}[!t]
    \centering
    \includegraphics[width=1\linewidth]{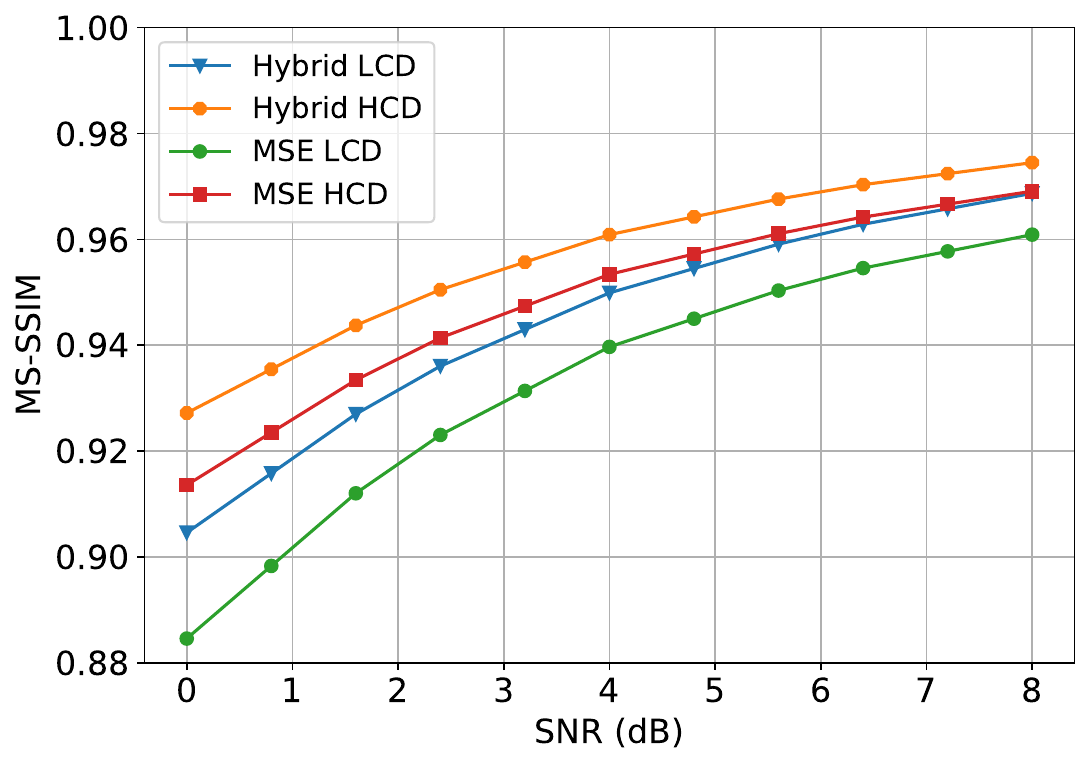}
    \caption{The comparison in MS-SSIM between the network trained with the MSE and Hybrid losses.}
    \label{MSSSIMHybridandMSE}
\end{figure}

As shown in Fig.~\ref{MSSSIMHybridandMSE}, we compare the performance of the network trained with MSE and hybrid loss under a number of channel conditions. Overall, the reconstructed images from the network trained with hybrid loss obtain higher MS-SSIM values than those from the network with MSE loss. In particular, the MS-SSIM gap between the two LCDs is around 0.02 when the SNR value is 0, and this gap gets smaller when the SNR is high, which indicates both models can perform well in a good environment. A similar result is acquired for the HCDs; the value difference is around 0.096 at SNR=0 and decreases if the noise from the physical channel is lower. The improvements in the MS-SSIM results are understandable due to the presence of MS-SSIM loss in our objective function. 
\begin{figure}[!t]
    \centering
    \includegraphics[width=1\linewidth]{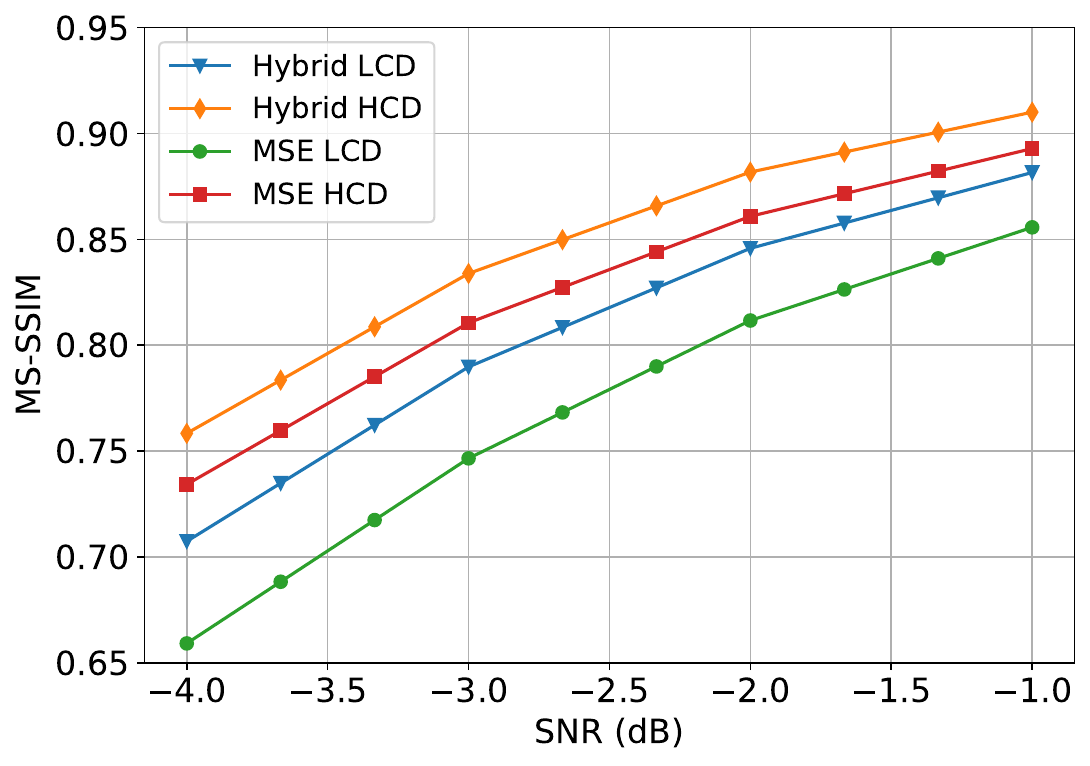}
    \caption{The performance difference of two training losses under harsh and inexperienced physic channel conditions.}
    \label{msssimharshcondition}
\end{figure}
\begin{figure*}[t!]
    \centering
    \includegraphics[width=0.83\textwidth]{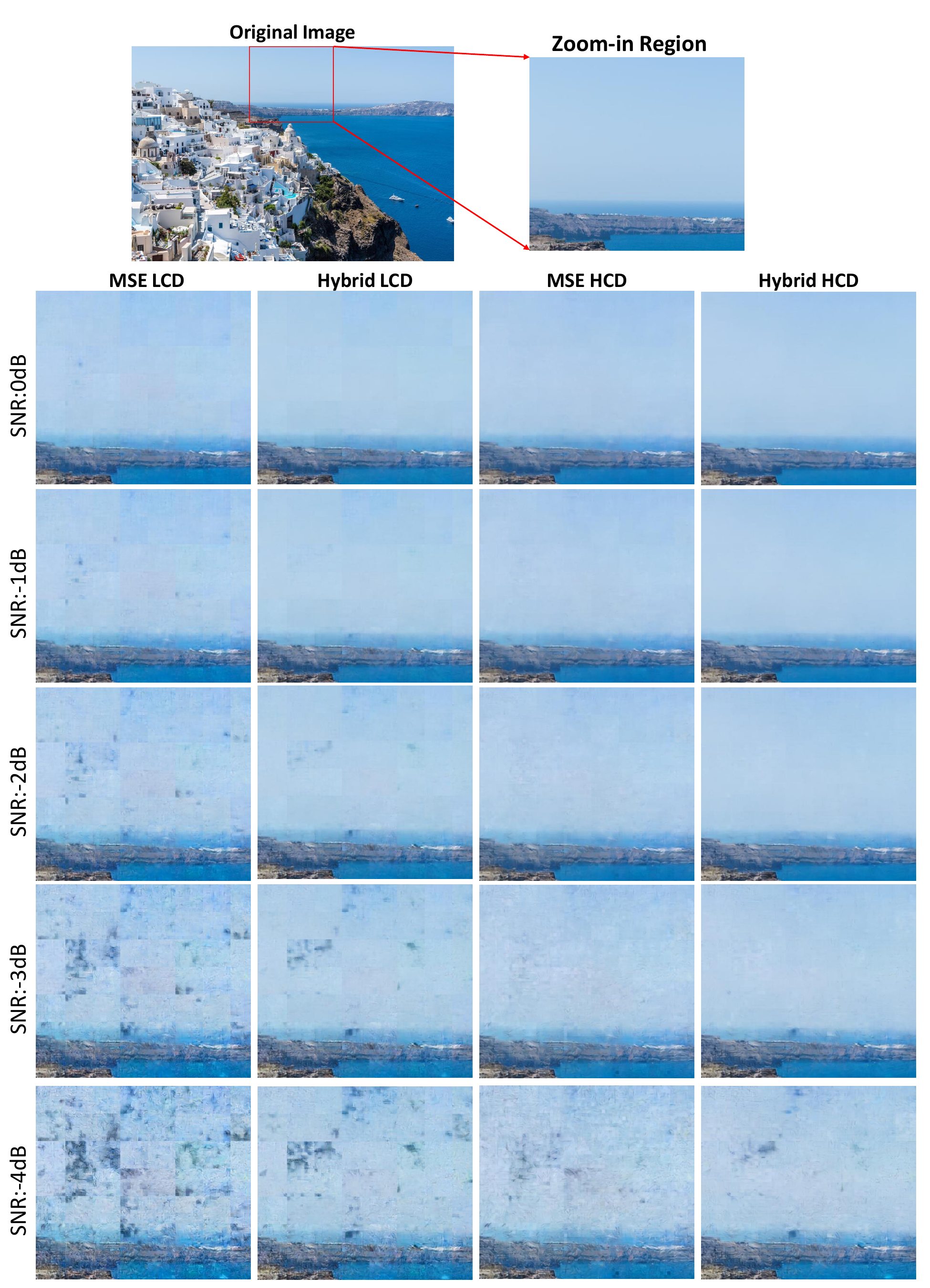}
    \caption{The reconstructed images under harsh and inexperienced physical channel conditions.}
    \label{Reconstructedimage}
\end{figure*}
\begin{table*}[t]
\centering
\caption{the psnr performance of the adjscc with two losses}
\label{PSNRADJSCC}
\renewcommand{\arraystretch}{1.1}
\begin{tabular}{|c|cc|cc|}
\hline
          & \multicolumn{2}{c|}{ADJSCC High}                                  & \multicolumn{2}{c|}{ADJSCC Low}                                   \\ \hline
SNR value & \multicolumn{1}{c|}{MSE}                  & Hybrid                & \multicolumn{1}{c|}{MSE}                  & Hybrid                \\ \hline
0 dB      & \multicolumn{1}{c|}{27.4740 $\pm$ 0.0017} & 27.4863 $\pm$ 0.0023  & \multicolumn{1}{c|}{27.1771 $\pm$ 0.0017} & 27.1764  $\pm$ 0.0042 \\ \hline
2 dB      & \multicolumn{1}{c|}{28.5634 $\pm$ 0.0017} & 28.5776 $\pm$ 0.0019  & \multicolumn{1}{c|}{28.3311 $\pm$ 0.0239} & 28.3299 $\pm$ 0.0036  \\ \hline
4 dB      & \multicolumn{1}{c|}{29.5056 $\pm$ 0.0017} & 29.5155  $\pm$ 0.0018 & \multicolumn{1}{c|}{29.3151 $\pm$ 0.0009} & 29.2923  $\pm$ 0.0034 \\ \hline
6 dB      & \multicolumn{1}{c|}{30.2773 $\pm$ 0.0016}  & 30.2889  $\pm$ 0.0016 & \multicolumn{1}{c|}{30.1268 $\pm$ 0.0008}  & 30.0796  $\pm$ 0.0030 \\ \hline
8 dB      & \multicolumn{1}{c|}{30.8839 $\pm$ 0.0014}  &30.8872 $\pm$ 0.0016   & \multicolumn{1}{c|}{30.7727 $\pm$ 0.0006}  & 30.6819 $\pm$ 0.0028   \\ \hline
\end{tabular}
\end{table*}
\begin{figure}[!t]
    \centering
    \includegraphics[width=1\linewidth]{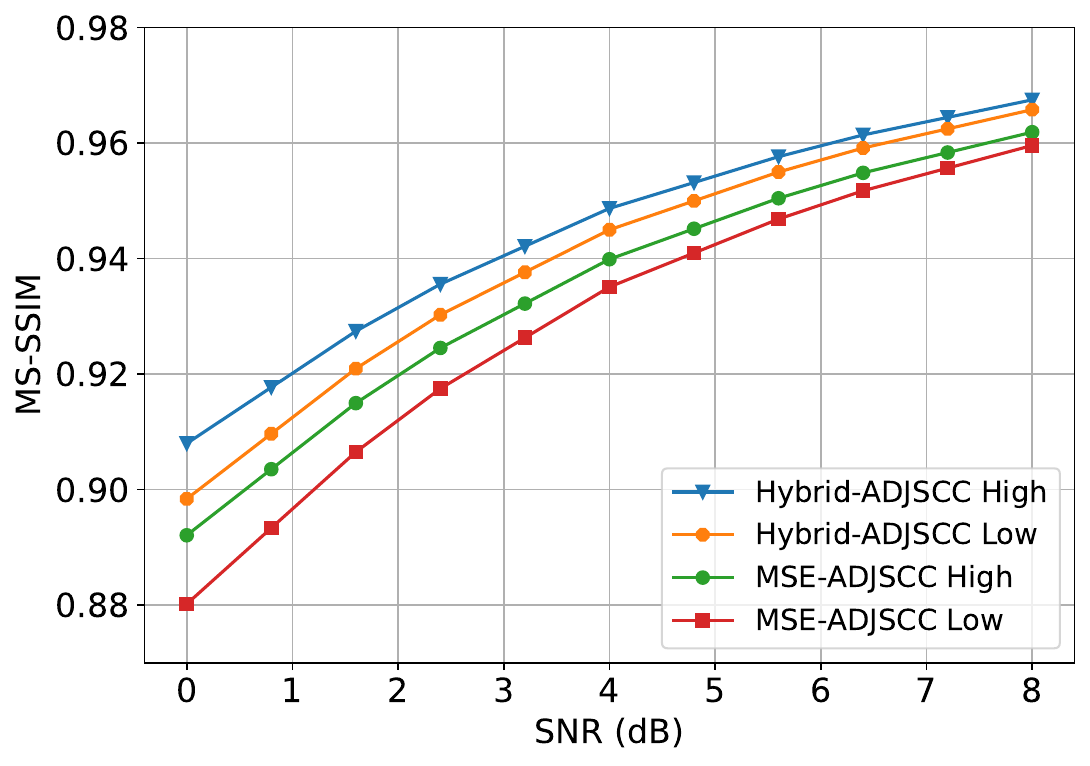}
    \caption{The performance comparison of MSE and Hybrid loss for the ADJSCC model.}
    \label{MSSSIMMSEandhybridADJSCC}
\end{figure}
\begin{table*}[t]
\centering
\caption{the performance change with different CBR under various channel conditions.}
\renewcommand{\arraystretch}{1.1}
\label{CBR}
\begin{tabular}{|cc|ccc|ccc|}
\hline
\multicolumn{2}{|c|}{}                                & \multicolumn{3}{c|}{LCD}                                                                                     & \multicolumn{3}{c|}{HCD}                                                                                     \\ \hline
\multicolumn{1}{|c|}{SNR}             &   \diagbox[width=6em]{Metric}{CBR}     & \multicolumn{1}{c|}{3/64}                 & \multicolumn{1}{c|}{4/64}                 & 5/64                 & \multicolumn{1}{c|}{3/64}                 & \multicolumn{1}{c|}{4/64}                 & 5/64                 \\ \hline
\multicolumn{1}{|c|}{\multirow{2}{*}{0 dB}} & PSNR    & \multicolumn{1}{c|}{26.6496 $\pm$ 0.0027} & \multicolumn{1}{c|}{27.3811 $\pm$ 0.0044} & 27.8956 $\pm$ 0.0003 & \multicolumn{1}{c|}{27.8727 $\pm$ 0.0025} & \multicolumn{1}{c|}{28.6462 $\pm$ 0.0022} & 29.1580 $\pm$ 0.0040 \\ \cline{2-8} 
\multicolumn{1}{|c|}{}                      & MS-SSIM & \multicolumn{1}{c|}{0.8890 $\pm$ 0.0001}  & \multicolumn{1}{c|}{0.9096 $\pm$ 0.0001}  & 0.9223               & \multicolumn{1}{c|}{0.9149 $\pm$ 0.0001}  & \multicolumn{1}{c|}{0.9318}               & 0.9416               \\ \hline
\multicolumn{1}{|c|}{\multirow{2}{*}{2 dB}} & PSNR    & \multicolumn{1}{c|}{27.7758 $\pm$ 0.0042} & \multicolumn{1}{c|}{28.3966 $\pm$ 0.0050} & 28.8191 $\pm$ 0.0014 & \multicolumn{1}{c|}{29.0682 $\pm$ 0.0006} & \multicolumn{1}{c|}{29.6853 $\pm$ 0.0026} & 30.0902 $\pm$ 0.0025 \\ \cline{2-8} 
\multicolumn{1}{|c|}{}                      & MS-SSIM & \multicolumn{1}{c|}{0.9215 $\pm$ 0.0001}  & \multicolumn{1}{c|}{0.9348 $\pm$ 0.0001}  & 0.9431 $\pm$ 0.0001  & \multicolumn{1}{c|}{0.9411 $\pm$ 0.0001}  & \multicolumn{1}{c|}{0.9509}               & 0.9567               \\ \hline
\multicolumn{1}{|c|}{\multirow{2}{*}{4 dB}} & PSNR    & \multicolumn{1}{c|}{28.7059 $\pm$ 0.0028} & \multicolumn{1}{c|}{29.2291 $\pm$ 0.0042} & 29.5654 $\pm$ 0.0015 & \multicolumn{1}{c|}{30.0122 $\pm$ 0.0011} & \multicolumn{1}{c|}{30.5120 $\pm$ 0.0030} & 30.8431 $\pm$ 0.0019 \\ \cline{2-8} 
\multicolumn{1}{|c|}{}                      & MS-SSIM & \multicolumn{1}{c|}{0.9416 $\pm$ 0.0001}  & \multicolumn{1}{c|}{0.9506 $\pm$ 0.0001}  & 0.9560               & \multicolumn{1}{c|}{0.9559}               & \multicolumn{1}{c|}{0.9624}               & 0.9662               \\ \hline
\multicolumn{1}{|c|}{\multirow{2}{*}{6 dB}} & PSNR    & \multicolumn{1}{c|}{29.4620 $\pm$ 0.0014} & \multicolumn{1}{c|}{29.8815 $\pm$ 0.0027} & 30.1331 $\pm$ 0.0009 & \multicolumn{1}{c|}{30.7586 $\pm$ 0.0010} & \multicolumn{1}{c|}{31.1535 $\pm$ 0.0024} & 31.4172 $\pm$ 0.0018 \\ \cline{2-8} 
\multicolumn{1}{|c|}{}                      & MS-SSIM & \multicolumn{1}{c|}{0.9551}               & \multicolumn{1}{c|}{0.9609}               & 0.9645               & \multicolumn{1}{c|}{0.9656}               & \multicolumn{1}{c|}{0.9699}               & 0.9724               \\ \hline
\multicolumn{1}{|c|}{\multirow{2}{*}{8 dB}} & PSNR    & \multicolumn{1}{c|}{30.0010 $\pm$ 0.0066} & \multicolumn{1}{c|}{30.3356 $\pm$ 0.0092} & 30.5237 $\pm$ 0.0060 & \multicolumn{1}{c|}{31.3133 $\pm$ 0.0030} & \multicolumn{1}{c|}{31.6197 $\pm$ 0.0041} & 31.8254 $\pm$ 0.0018 \\ \cline{2-8} 
\multicolumn{1}{|c|}{}                      & MS-SSIM & \multicolumn{1}{c|}{0.9634}               & \multicolumn{1}{c|}{0.9673}               & 0.9697               & \multicolumn{1}{c|}{0.9717}               & \multicolumn{1}{c|}{0.9746}               & 0.9764               \\ \hline
\end{tabular}
\end{table*}

As mentioned in Section \ref{traningdetail}, the models are trained under a range of SNR from 1 to 7. The models still perform well while being tested with an SNR value inside or slightly outside of this range. However, we want to examine the performance under worse conditions that the models have not experienced before as shown in the table.~\ref{PSNRunderharshenvironment} and  Fig.~\ref{msssimharshcondition}. In terms of PSNR values, the proposed loss slightly outperforms MSE in all channel conditions for both low-computing and high-computing decoders. One notable point is the significant improvement in the MS-SSIM metric, as shown in Fig.~\ref{msssimharshcondition}, where a lower SNR value results in a wider performance gap. 

Besides these evaluation metrics, we investigate the visual quality of the reconstructed image under these harsh conditions in Fig.~\ref{Reconstructedimage}. When the SNR value is at 0 dB, there is not much difference in visual details between the reconstruction from the networks trained with MSE and hybrid loss compared to the original image. At this SNR level, we only can recognize the visual difference between the reconstructed image by LCD and the HCD where the pixels for the sky in both MSE LCD and Hybrid LCD are not as smooth compared the HCD. When the SNR decreases to -1, the reconstructed image by MSE LCD has some dark blue noises in the sky, while the sky in the hybrid LCD is still seamless. Only when the SNR value drops to below -2 dB, noise starts to appear in the reconstructed image of the LCD hybrid. Both models produce low-quality images due to the effect of noise level at -4 dB, however, the visual quality of the proposed loss is still better compared to the MSE loss. The HCDs are more resistant to channel noise than the LCD. The noise effects only appear notably when the SNR is below -2 dB. If we carefully examine the image reconstructed by the HCD trained with MSE when the SNR value is at -2, it contains a large number of splotchy artifacts distributed over the sky, and they become more clear when the channel quality degrades to -3 or -4 dB. The image produced by HCD trained with the hybrid loss does not contain any splotchy artifacts when the SNR value is greater than -2 dB and the spots only start to appear at -3 dB. From the the results of MS-SSIM, PSNR metrics, and human visual examinations, we can conclude that the model trained with hybrid loss is more stable and produces visually pleasing images than MSE loss, especially in harsh environments which the model has not encountered in training process.

In addition, we rebuild the ADJSCC proposed in \cite{xu2021wireless} to prove the general efficiency of the hybrid loss. The encoder contains six feature learning modules and five attention feature (AF) modules, the high-computing decoder has the same architecture, while the low-computing decoder only contains four feature learning modules and three AF modules. To differentiate these high-computing and low-computing decoders apart from the decoders in the above case, we call them ADJSCC high and low. In these simulations, we focus on the improvement made by our hybrid loss, and not on the design of the DL network. Therefore, we keep the network architecture the same for all the simulations, and only change the objective function. We also implement the broadcast case. As shown in Table~\ref{PSNRADJSCC}, the images produced by the hybrid loss achieve competitive PSNR results compared to the MSE loss for both low and high computing decoders, while obtaining significant improvement in the MS-SSIM metric. Figure~\ref{MSSSIMMSEandhybridADJSCC} shows the trend of the MS-SSIM value under different channel environments, which is similar to the proposed system. The network trained with hybrid loss outperforms the one trained with MSE by a large margin. Even the hybrid low-computing decoder achieves higher performance than the MSE high computing one.
% \usepackage{multirow}
% \usepackage{multirow}

\begin{figure*}[t!]
    \centering
    \includegraphics[width=0.95\textwidth]{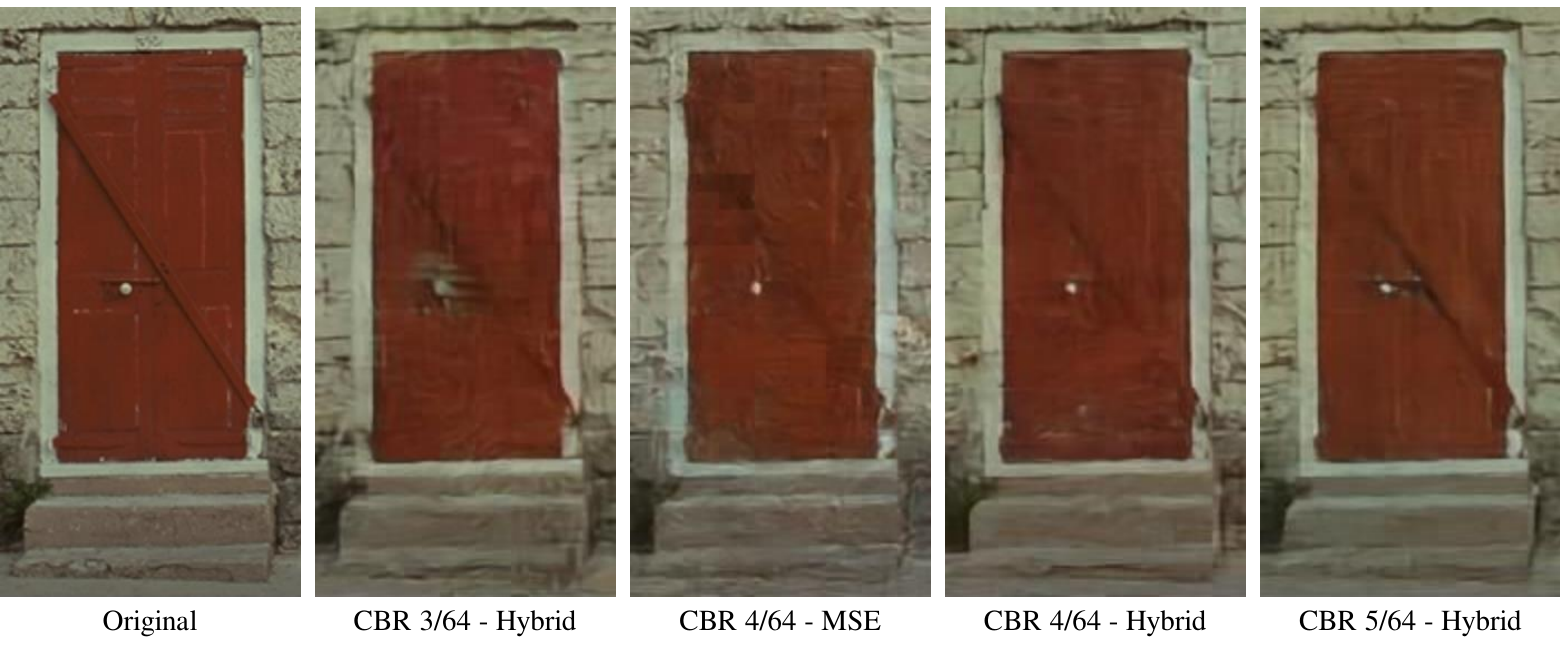}
    \caption{The reconstructed images with different compression ratios at -1dB and their corresponding PSNR and MS-SSIM: [23.578; 0.831], [24.113; 0.834], [24.315; 0.860], [24.868; 0.883].}
    \label{Kodak}
\end{figure*}

\subsection{The performance of the proposed dynamic compression ratio}

In the previous sub-sections, we fixed the CBR to 4/64 and optimized the network with this CBR alone in order to highlight the performance gain from the proposed loss and framework. Thus, there is only one FC layer inside the compression ratio module. However, in this sub-section, we want to evaluate the efficiency of the dynamic CBR property of the channel encoder, so there will be three FC layers within the compression ratio module. Table~\ref{CBR} shows the mean and variance values for PSNR and MS-SSIM metrics over the validation images of the DIV2K dataset. As we can see in the table, the higher the CBR, the higher the value of the evaluating metrics. To be more specific, when the SNR is fixed at 0 dB, the high computing decoder achieves the average PSNR up to 29.1580 and 0.9416 for the MS-SSIM metric when the CBR is 5/64. These metric values decrease to 27.8727 and 0.9149 when the CBR decreases to 3/64. The decrease in performance is acceptable if we realize there is a considerable difference in size between these two ratios. To be accurate, the output length of the former compression ratio is 66.6\% longer than the latter. The performance of the low-computing decoder experiences the same trend as the high computing one for different CBRs. We observed that reconstructed images by the HCD have 1.2 higher PSNR value than images from the LCD for in every CBR value at 0dB. In the meantime, the MS-SSIM value difference between the HCD and the LCD are 0.0259, 0.0222, and 0.0193 for the CBR: 3/64, 4/64, and 5/65 at SNR=0, respectively. These value gaps decrease as the SNR increases. The gap reduction does not mean the performance of any decoder degrades or improves compared to the other, simply because they are saturated when approaching the maximum MS-SSIM value.

% or


In addition to the above results, we validate the performance of the compression ratio module with the hybrid loss using another dataset, which is the Kodak dataset. The experiments still obtain promising results for the PSNR and MS-SSIM metrics. In Fig.~\ref{Kodak}, we show the reconstructed images by the high computing decoder with an SNR value of -1 dB in different compression ratios and compare them with the reconstructed image by the network trained with the MSE loss. First, we compare the image produced by MSE loss and the image produced by hybrid loss with the same CBR. The hybrid loss outperforms the MSE loss in both PSNR and MS-SSIM metrics. In terms of image visual quality, we can see that our proposed method achieves better smoothness compared to the MSE case in the zoomed-in mode, where a couple of pixel regions suddenly change their color from red to black and reverse on the door. Moreover, there is no abrupt shift in color for our case, which pleases the human visual system. We additionally demonstrate the image produced by the same neural network but with different compression ratios. The increase in the CBR leads to an increase in image quality, both in evaluation metrics and the visual perspective. As we can see in the figure, the doorknob and the handle of the door are blurry when the CBR is at 3/64, become clearer at 4/64, and  most visible at 5/64. These results not only validate the efficiency of the dynamic compression module of the channel encoder but also the loss objective.

\section{Conclusion}\label{Conclusion}

In this work, we proposed a realistic study of the semantic communication system by considering the multi-user scenario while also assuming there is a gap in the computing capacity of users. Furthermore, we proposed the targeting embedding vector at the transmitter to extract the semantic feature that is particularly suited for one type of receiver, preventing high-quality reconstructions for another type of receiver. In addition, we proposed a hybrid loss, which outperformed the MSE loss in evaluating metrics and generating images that satisfy the human visual system. Last but not least, the channel encoder of our semantic communication system can dynamically adjust the transmission length of the output to adapt to the network demand and also eliminate the channel noise with the SNR feedback. Throughout this paper, we highlighted the significance of considering the varying computing capacities of DL models belonging to multi-users, even though it may introduce training instability. Another key finding from this paper is the effectiveness of using MS-SSIM and AE losses in generating high-quality images with impressive visual fidelity.

% In this work, we only focused on image transmitting, however, modern applications nowadays already require the transmission of multi-modality data. Therefore, we will expand this work to transmit multi-modality in the near future.
\appendices

\ifCLASSOPTIONcaptionsoff
  \newpage
\fi

\bibliographystyle{IEEEtran}
\bibliography{mybib}
\end{document}